# Observation of electron excitation into silicon conduction band by slow-ion surface neutralization

S Shchemelinin and A Breskin

*Department of Astrophysics and Particle Physics, Weizmann Institute of Science, 7610001 Rehovot, Israel*

E-mails: sergei.shchemelinin@weizmann.ac.il,  amos.breskin@weizmann.ac.il.

## Abstract

Bare reverse biased silicon photodiodes were exposed to 3eV $He^+$, $Ne^+$, $Ar^+$, $N^{2+}$, $N^+$ and $H_2O^+$ ions. In all cases an increase of the reverse current through the diode was observed. This effect and its dependence on the ionization energy of the incident ions and on other factors are qualitatively explained in the framework of Auger-type surface neutralization theory. Amplification of the ion-induced charge was observed with an avalanche photodiode under high applied bias. The observed effect can be considered as ion-induced internal potential electron emission into the conduction band of silicon. To the best of our knowledge, no experimental evidence of such effect was previously reported. Possible applications are discussed.

Keywords: auger ion neutralization, ion interaction with surfaces, ion-induced electron emission, detection of slow ions, solid state detectors, ion deceleration.

## 1. Introduction

When an ion of very low kinetic energy is neutralized at a solid surface, its potential energy (corresponding to the ionization energy of the parent atom or molecule) is released; most of this energy is transferred to electrons of the solid via Auger neutralization [1]. A well-known consequence of such a process is the ejection of electrons from the solid into vacuum, referred to as *potential* ion-electron emission; this term is derived from the fact that the process is due to the ion's potential energy, while its kinetic energy does not play here any significant role. This phenomenon was studied both theoretically and experimentally since the 1930s [2-4]. A detailed theoretical study of ion Auger neutralization at silicon and germanium surfaces followed by electron emission into vacuum was made by Hagstrum [1]. He suggested that, in some cases, electrons may only be excited into the conduction band via Auger neutralization without their ejection to vacuum. This prediction was out of the mainstream of his research that was rather focused on electron emission – as a tool for surface studies (e.g. in ion-neutralization spectroscopy [5]). Nevertheless, charge carriers remaining in the conduction band should affect the solid's conductivity properties. Such effect could for example pave the way to novel detector concepts of slow ions and neutral metastable atoms and molecules. In this letter we report on experimental observations of an increase in silicon conductivity due to neutralization of low-energy ions at its surface. No experimental evidence to that has been reported, to the best our knowledge, in literature.

## 2. Experiments

### 2.1. Samples selection

The experiments performed required Si samples with low concentration of charge carriers in its conduction band – to enhance sensitivity to the desired effect. Such low concentration can be obtained in reverse-biased diodes. Therefore, in this first study, we opted for commercially-available bare photodiodes, with the thinnest possible dead-layers. The latter condition is a prerequisite for the efficient detection of electrons excited to the conduction band by surface neutralization. In the diode selection process we relied on the UV cutoff wavelength, bearing in mind that the thinner the dead-layer the shorter is the UV cutoff wavelength [6]. The chosen photodiodes were Hamamatsu S9904 (referred below as PD) and Advanced Photonix APD 197-70-73-520 (referred below as APD); they were slightly modified for us by the manufacturers: Hamamatsu PD was manufactured without its usual central hole and Advanced Photonix APD – without anti-reflection coating[1].

### 2.2. Experimental procedure

The selected photodiodes were irradiated with 3eV positive ions: $He^+$, $Ne^+$, $Ar^+$, $N_2^+$, $N^+$ and $H_2O^+$. A conceptual scheme of the performed experiment and an example of a typical result are shown in figure 1. The ions were generated by a Collutron[2] Ion Gun G1 coupled to a home-made ion decelerator – drscribed in more detail in secttion 2.3. For proper dark current subtraction, a beam-chopper was implemented, periodically switching the ion beam on and off the target. The beam was directed onto the sensitive surface of the reversed-biased photodiode; the currents $I_1$ and $I_2$ from its two sides were measured, as depicted in figure 1(a). An example of the measured currents is shown in figure 1(b). $I_2$ depends on the concentration of the charge carriers in the depleted region and can be considered as a sum of the dark current $I_{dark}$, and the expected ion-induced conductivity current $I_{cond}$. From the balance of the incoming ions current $I_{ion}$, $I_1$ and $I_2$ one can conclude that:

$$I_1 = I_{dark} + I_{ion} + I_{cond} \tag{1}$$

---

[1] See the prototype parameters correspondingly at www.datasheetarchive.com/S9904/Datasheets-IS69/DSAH00186682.html and www.chipfind.net/datasheet/photonix/1977073520.htm

[2] www.colutron.com

The observed drift of $I_2$ in figure 1 (b) indicates a small variation of $I_{dark}$ during the measurements (~0.05% per minute), that did not affect the results. $I_{ion}$ measured at the sample was typically in the range of 0.3 – 3 nA.

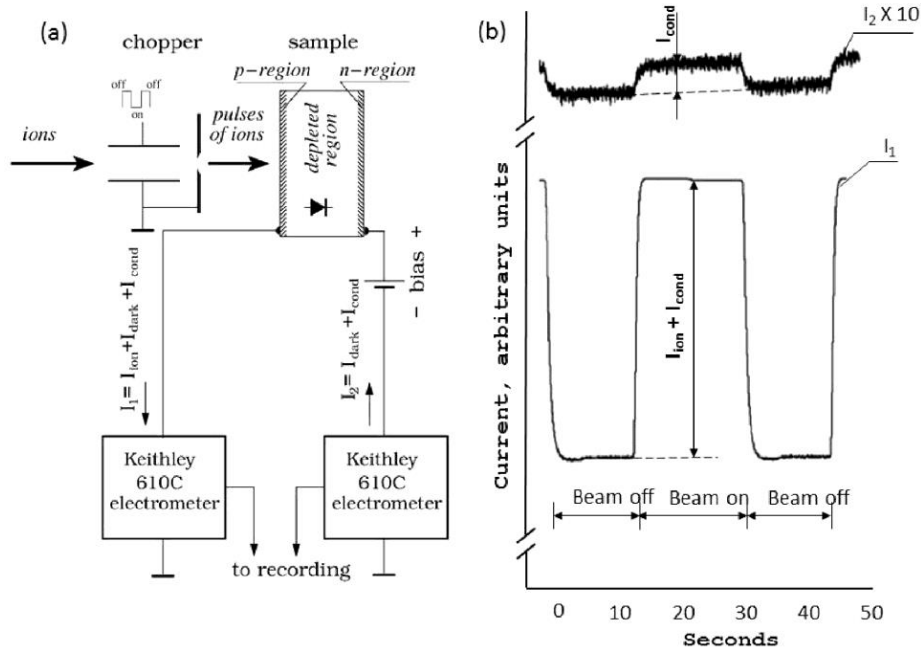

**Figure 1. Scheme of the measurements with a chopped ion beam (a) and an example of the results (b), averaged over 163 records. $I_{ion}$ is the current of incoming ions, $I_{dark}$ is the background current through the diode, and $I_{cond}$ is the current of the ion-induced conductivity.**

### 2.3. Slow-ions generation

A scheme of the experimental setup providing a beam of mass-selected slow ions, of kinetic energies down to a few eV, is shown in figure 2. The ion beam was generated by a modified by us Colutron ion-gun system G1. The beam from the ion source, composed of various ion species, was accelerated up to about 1 keV and entered a Wien velocity filter – tuned on a given ion mass. The mass-selected ions were directed onto the surface of the investigated sample via a stack of focusing and decelerating electrodes; the sample was kept at ground potential. In this setup, a potential $U_0$ of a few volts (above ground) was applied to the ion source, whereas the potentials at the accelerator's cathode, the velocity filter and all the parts preceding the end station were kept at the accelerating potential $U_{acc}$ with respect to the ion source (-1kV in this work). In that way the kinetic energy of $q$-charged ions in the velocity filter was 1 keV, regardless of $U_0$; after deceleration, their energy $E_0=q*U_0$ at the grounded sample did not depend on $U_{acc}$. This allowed for varying $E_0$ without re-tuning the velocity filter.

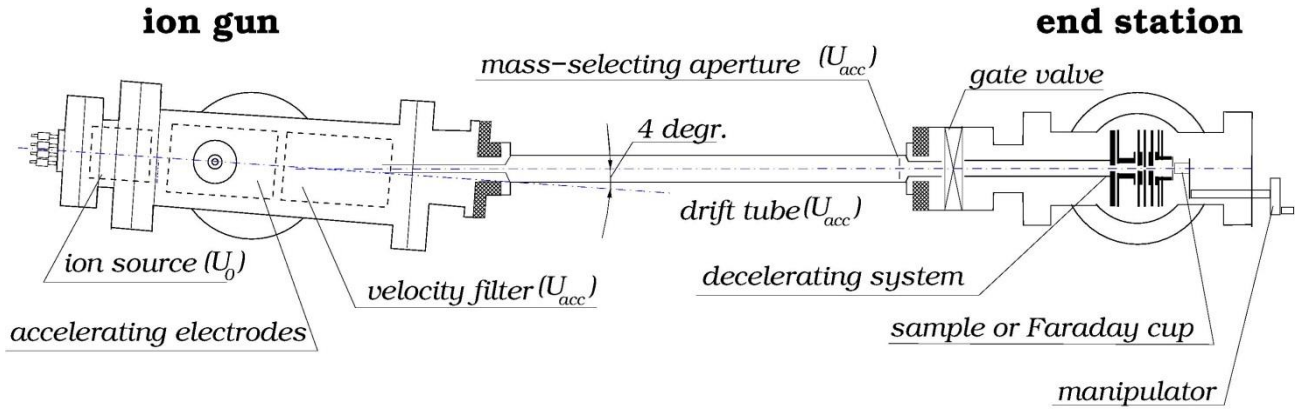

**Figure 2. A general scheme of the ion acceleration-deceleration setup. The sample at the end station is irradiated by slow mass-selected ions of energy $E_0=qU_0$. The potentials at different parts are given in parenthesis (See text).**

In this setup the velocity filter operated in an "angle-shift" mode, in which the selected ions at its output were deflected off the acceleration-stage axis by 4 degrees (figure 2). This significantly reduced the amount of residual light from the ion source as well as the neutral fraction of the beam reaching the sample; both would cause a significant background current though the sample in a usual zero-degree aligned system. For improving the mass resolution, the setup included a drift tube, having a mass-selecting aperture at its end. The distance between the sample and the deceleration system was dictated by the short focus of the latter. The vacuum level in the end station was maintained below $3 \cdot 10^{-7}$ mbar during experiments.

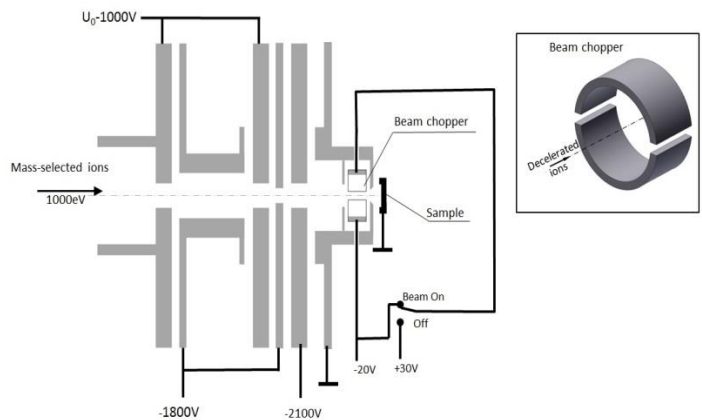

**Figure 3. The decelerating electrode system. Mass-selected ions enter the electrode system with their kinetic energy of 1000 eV; after further acceleration and deceleration they are focused onto the sample – kept at ground potential. Their final energy is defined by the ion-source potential $U_0$ (see text).**

The deceleration system with the beam chopper and the sample location are shown in figure 3. The electrode assembly was manufactured by us according to the Colutron Research Corporation recommendations [7]. The chopper was designed as a cylindrical electrode cut into two halves along its axis (see insert in figure 3). In a "beam on" situation, the two halves were kept under the same potential, with the chopper operating as an electrode focusing the beam onto the sample; in the "beam off" position, a potential difference was applied between the two parts, with the ions getting deflected off the sample. The chopper was placed close to the sample, to minimize the effects of undesirable "background" currents (see above).

## 3. Results and discussion

Two types of experiments were performed: (1) with both diodes in *"charge-collection"* mode (without gain in the APD) and (2) with the APD in a *"multiplication"* mode. In the charge-collection mode (1) the measurements were performed according to the scheme in figure 1(a). The applied reverse bias was about 10 mV for the PD and about 1 V for the APD; at this applied voltage the APD operated as a standard photodiode without avalanche gain. In all cases, a pronounced increase of the reverse diode current was observed under ion-beam impact on the diode surface, (see figure 1(b)). The ratio $\alpha = I_{cond}/I_{ion}$ is defined as the value characterizing the response of the diode to incident ions. It represents the number of collected secondary electrons (in the conduction band) per single ion neutralized at the surface. The measured values of $\alpha$ are presented in figure 4 versus the ionization energy $E_i$, carried by the different incident-ion species.

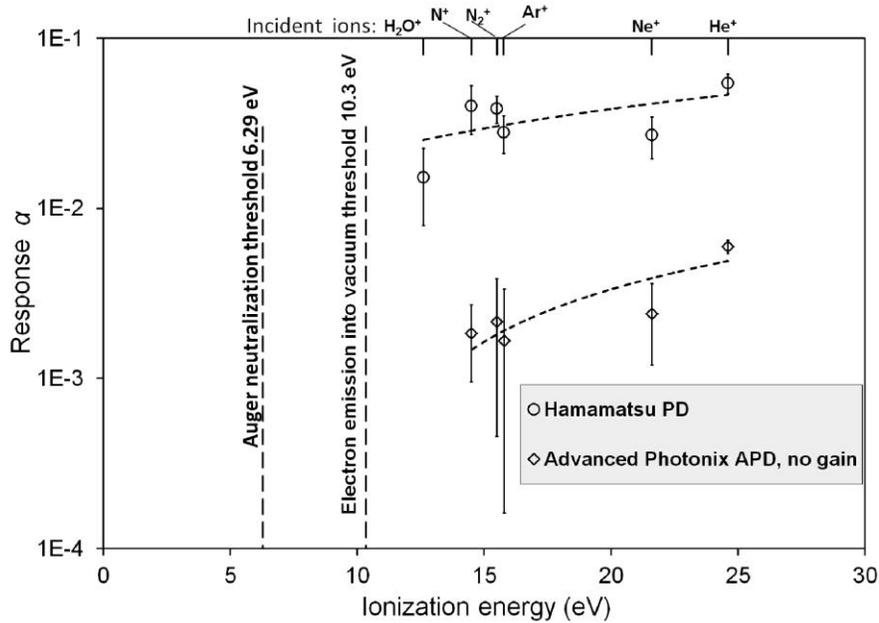

**Figure 4 . Response of the Hamamatsu PD and Advanced Photonix APD (reversed-bias 1V) to irradiation with 3eV ions of different ion species, vs their ionization energy $E_i$ (shown on top for the various species). The values of $\alpha = I_{cond}/I_{ion}$ are shown for reverse conductivity current $I_{cond}$ trough the diode, induced by incident ion current Iion. The dashed curves are linear fits. The vertical dashed lines show the energy thresholds for the Auger neutralization and electron emission into vacuum, calculated according to Hagstrum theory[1].**

The error bars in figure 4 are of ±10% for He$^+$ and significantly larger for other incident-ion species, of lower $E_i$ values. The origin of the very large observed fluctuations for the heavier ions is yet unknown – requiring further investigations; their dependence on $E_i$ might be, for example, due to surface impurities. Nevertheless, in spite of the observed fluctuations, one can clearly observe in these first results two notable features: (1) significantly (about tenfold) larger values of $\alpha$ for the PD than for the APD and (2) an increase, in both samples, of $\alpha$ with $E_i$. Both features can be explained in the framework of an Auger-type neutralization theory. The measured value $\alpha$ can be considered as the product of two factors: the efficiency of creating a charge carrier in the ion-neutralization process and the probability of the charge-carrier survival before its collection at the diode's electrodes (the dead-layer effect). To explain (1) it is suggested that the PD has a thinner dead-layer compared to the APD, as the former was specially designed for slow-electron detection (secondary electrons in electron microscopy), which, likely to our experiments, requires a thing dead-layer. To explain (2), one can assume that the survival probability of a charge-carrier during its drift across the dead-layer increases with its initial energy, which in turn increases with the ion's $E_i$. An analogous increase with $E_i$ was clearly observed for ion-electron emission into vacuum in many experiments [4].

In multiplication mode the potential of 1880 V applied to the APD yielded gain of 66. In this case, the high applied voltage prevented us from measuring $I_2$ according to the scheme in figure 1(a). Instead, only $I_1$ was measured, but *twice*, at two values of the potential applied to the *n*-side of the APD. For the multiplication mode, equation (1) should be replaced by:

$$I_1 = I_{dark} + I_{ion} + I_{cond} \times G \qquad (2)$$

where $G$ is the APD gain. The net ion-related current $I_G$ through the APD, having this gain, can be obtained from the beam-chopped step in $I_1$ seen in figure 1(b). It can be written as

$$I_G = I_1 - I_{dark} = I_{ion}(1 + \alpha G) \qquad (3)$$

with $\alpha = I_{cond}/I_{ion}$. Although $I_{cond}$ is the ion-induced conductivity current *before* APD amplification, $\alpha$ (not depending on G explicitly) may still depend on the APD voltage (see below). For a low APD voltage, with $G=1$, equation (3) can be rewritten as

$$I_0 = I_{ion}(1 + \alpha_0) \qquad (4)$$

with $I_0$ and $\alpha_0$ representing the corresponding values of $I_G$ and $\alpha$ at a low APD voltage, when $G=1$. Using (3) and (4), $\alpha$ can be derived from measured $I_G$ and $I_0$:

$$\alpha = \frac{I_G - I_0}{I_0 G}(1+\alpha_0) + \frac{\alpha_0}{G} \qquad (5)$$

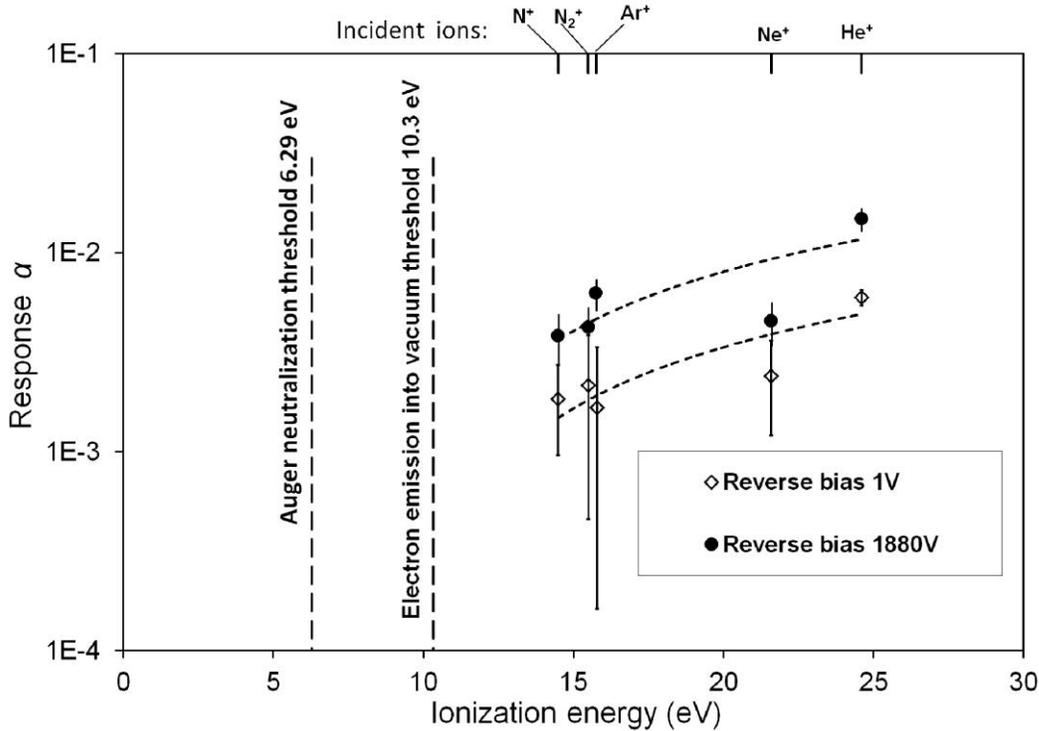

**Figure 5. . Comparison of the Advanced Photonix APD response $\alpha = I_{cond}/I_{ion}$ to 3eV incident ions at applied potentials of 1 V ($G=1$) and 1880 V ($G=66$). Note that in both cases $I_{cond}$ is the initial conductivity current trough the diode, induced by the surface neutralization before the APD amplification. The Ei values of the different ion species are shown on top. The dashed curves are linear fits.**

The measurements of $I_G$ and $I_0$ were done at the corresponding applied voltages of 1880 V (gain 66) and 1 V (gain 1); the values of $\alpha_0$, derived from figure 4, were introduced into relation (5).

The obtained values of $\alpha$ vs $E_i$ in multiplication mode are depicted in figure 5, together with that of $\alpha_0$. One can see a strong correlation between the two graphs, with a ~2.5 fold increased response under the high applied voltage. Hence the reverse conductivity current $I_{cond}$ induced by the surface neutralization, *before* the APD amplification, increases under high applied potential. In this situation, the latter is sufficiently high to establish in the diode an electric field - of the same order as the diode's internal one (without external voltage). This additional field reduces the probability of electron-hole recombination in the near-surface dead-layer. The effect of increasing the collection efficiency of the carriers from the layer with increasing the electric field is known and has been analyzed theoretically [8] for UV-sensitive photodiodes.

In our experiments, the number of collected ion-induced charge carriers, per incident ion, reached at best 5% – for $He^+$ ions (Fig. 4). A possible cause could be a low collection efficiency of the charge carriers induced at the surface (a dead-layer effect). To support this hypothesis, we estimated the collection efficiency of the carriers independently, based on experimental data of the diode response to *electrons*. In this estimation we obtained the distribution of the collection efficiency of the carriers induced in the photodiode at various distances from the surface.

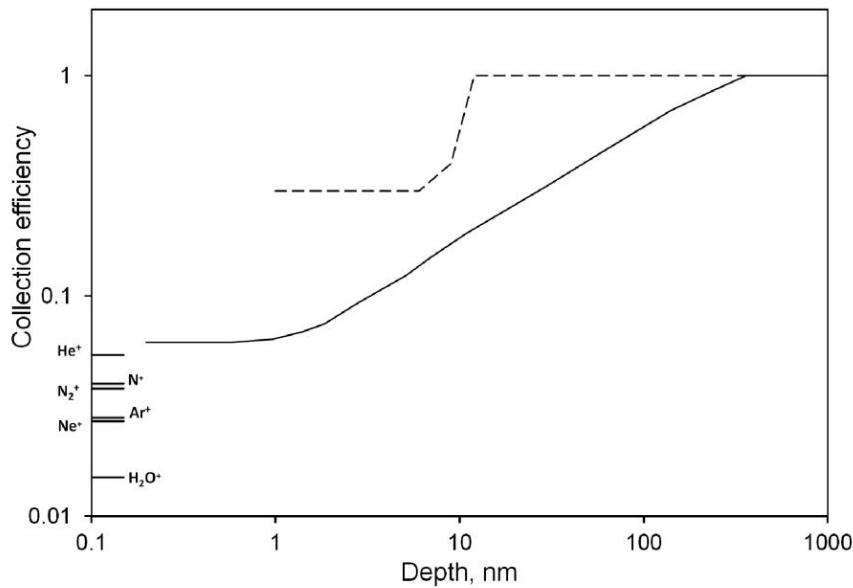

**Figure 6. Collection efficiency of the carriers induced in a photodiode at different depths. The solid line presents the fitted efficiency distribution for the Hamamatsu PD, which brings the calculations according to equation (6) into agreement with our measurements. The dashed curve is the efficiency distribution obtained in the same way for a delta-doped device [12]. The horizontal strokes on the vertical axis mark the responses of the Hamamatsu PD to various ion species.**

Both an additional experiment and CASINO[1] simulations were performed. The reverse-biased Hamamatsu PD was irradiated with electrons with energies in the range of 60 – 8700 eV and the "gain", defined here as the ratio of the current through the diode to the incident electron current, was measured; it varied from 0.54 at 60 eV to 1780 at 8700 eV. In the simulation, the "gain" was calculated as follows. The depth distribution of the energy deposited in silicon by an incident electron of given energy was simulated with CASINO. Assuming the density of the induced charge carriers being proportional to the deposited energy density (taking 3.71 eV per electron-hole pair), the obtained energy density distribution was converted to a depth distribution of the induced charge carriers *n(x, E)*, where *x* is the distance from the surface and *E* is the energy of the incident electron. The ratio of the current through the photodiode *I(E)* to the current of the incident electrons $I_{el}(E)$ was calculated by integration across the semiconductor:

$$\frac{I(E)}{I_{el}(E)} = \int n(x,E)f(x)dx \qquad (6)$$

with collection efficiency *f(x)* of the carriers induced at depth *x*. The latter was varied and fitted to bring the calculations into agreement with the measurements. The obtained *f(x)* is shown by the solid line in figure 6.

---

[1] CASINO - monte CArlo SImulation of electroNs inSOlids; can be used on-line, for example, http://www.gel.usherbrooke.ca/casino/

In our ion experiments, the carriers are induced at the surface only. Therefore the diode response $\alpha$ to ions (see figure 4) can be interpreted as the product of the probability of carriers' induction and their collection efficiency at *zero depth*. The values $\alpha$ for the Hamamatsu PD are marked in figure 6 on the vertical axis with horizontal strokes. Note that the described above efficiency estimation is not applicable to small depths; however, the efficiency at the surface can be estimated by extrapolating the solid curve in figure 4 to zero. From this one can conclude that the main factor limiting the response to ions is not the production probability of a charge carrier pair but rather its collection efficiency from the surface. In other words, even if the probability to induce an electron in the conduction band is close to 1, only a very small fraction of these electrons will succeed crossing the dead layer.

It should be noted that the diodes used in this work were not designed for our experiments; they were rather selected among commercially-available ones, designed for different applications. More promising for further experiments would be to have devises with optimized surfaces, like those obtained by delta doping [9, 10] with significantly reduced dead-layers in silicon [11]. For comparison, we estimated (in the same way as for the Hamamatsu PD) the collection efficiency distribution for a delta-doped device [12]; the measured response to incident electrons was taken from [12]. The estimated efficiency distribution is shown in figure 6 with the dashed line (the minimal depth of 1 nm is due to the lowest electron energy in [12], of 200 eV). From comparison of this efficiency with that of the Hamamatsu PD one can expect significantly higher response of delta-doped devices to incident slow ions of eV energies.

## 4. Conclusions

In this work we observed, for the first time, an increase in the reverse current of reverse-biased silicon photodiodes caused by neutralization of slow (3 eV) ions at their sensitive surfaces. The current increase depends on the incident-ion species and on the specific photodiode structure. It was observed that with the APD, the initial generation of ion-induced carriers (before charge multiplication) is significantly larger (2.5 fold) under high applied potential. Note that this increase (seen in figure 3) is not due to avalanche multiplication factor of the APD, but rather to the reduced recombination under the high electric field.

The reported effect and its observed feature form a self-consistent picture, which can be qualitatively explained in the frame of the Auger-type neutralization theory, taking into account the collection efficiency of the induced charge carriers.

This observed effect can be considered as *internal potential ion-electron emission* (into the conduction band). To the best our knowledge, no experimental evidences of such effect were reported in the literature.

Further studies are required, with dedicated surfaces, for deeper understanding of the observed phenomenon and for optimizing the yield of charge carrier production and collection; they are expected to assess the potential applicability of the effect to novel semiconductor structures capable of effective slow (down to sub-eV) ion detection. As the surface Auger neutralization is very similar to surface Auger de-excitation, such detectors would be possibly sensitive also to metastable neutral atoms or molecules. Naturally, the detection of ion-induced (or de-excitation-induced) charges would require, besides proper surfaces, the development of new concepts of internal charge amplification. While current slow-ion detectors usually require their pre-acceleration, requiring high vacuum, as described in [13], internal-emission detectors would operate even in a gas.

Among potential applications of slow-ion detectors are: ion-counting nanodosimetry [13] for track-structure studies in radiobiology, track-structure recording in astrophysics like in x-ray polarization studies [14] and rare-event experiments e.g. in directional dark-matter searches [15] and neutrinoless double-beta decay [16]. Detection of very slow ions is required also in ion-mobility spectroscopy [17] and ion detection in mass spectrometry.

## Acknowledgments

We are indebted to the late Dr. Y. Krichevets of Micro Components Ltd. (Migdal Haemek, Israel) for his assistance as well as to Dr. V. Dangendorf of PTB (Braunschweig, Germany) and Dr. M. Rappaport of the Weizmann Institute of Science, for fruitful discussions and useful comments. We acknowledge Dr. Erik Wåhlin of Colutron Research Corporation (Boulder, Colo. USA) for assistance in the ion decelerator design. A. Breskin is the W.P. Reuther Professor of Research in The Peaceful Use of Atomic Energy.